\documentclass[9pt,twocolumn,twoside]{opticajnl}

\journal{opticajournal} 
\usepackage{graphicx}
\usepackage{dcolumn}
\usepackage{lipsum}
\usepackage{bm}
\usepackage{tabu}
\usepackage{amsmath}
\usepackage[utf8]{inputenc}
\usepackage[T1]{fontenc}

\usepackage{color}

\newcommand{\bl}[1]{\textcolor{blue}{ #1}}

\usepackage{comment}
\usepackage{soul}

\newcommand{\hatt}[1]{\ensuremath\hat{{\mathbf #1}}}
\renewcommand{\vec}[1]{\ensuremath{\mathbf{#1}}}

\setboolean{shortarticle}{true}

\title{Generalized vector beams in elliptical degrees of freedom}
\author[1]{Xiao-Bo Hu}
\author[1]{Yu-Chen Chen}
\author[1]{Qi-Yao Yuan}
\author[2]{Valeria Rodriguez-Fajardo}
\author[3,$\dagger$]{\\Carmelo Rosales-Guzm\'an}
\author[4,*]{Benjamin Perez-Garcia}
\affil[1]{Key Laboratory of Optical Field Manipulation of Zhejiang Province, Department of Physics, Zhejiang Sci-Tech University, Hangzhou, 310018, China}
\affil[2]{Departmento de Física, Universidad Nacional de Colombia Sede Bogotá, Carrera 30 No. 45-03, Bogotá, 111321, Colombia}
\affil[3]{Centro de Investigaciones en Óptica, A.C., Loma del Bosque 115, Colonia Lomas del campestre, 37150 León, Gto., Mexico}
\affil[4]{Photonics and Mathematical Optics Group, Tecnologico de Monterrey, Monterrey 64849, Mexico}
\affil[$\dagger$]{Corresponding author: carmelorosalesg@cio.mx}
\affil[*]{Corresponding author: b.pegar@tec.mx}

\begin{abstract}
The strong coupling between the spatial and polarisation degrees of freedom (DoF) in vector modes enables a diverse array of exotic, inhomogeneous polarisation distributions through a non-separable superposition, which are conventionally generated in circular-cylindrical symmetry. Here, we theoretically and experimentally demonstrate a generalized class of vector modes specified in elliptical spatial coordinates and elliptical polarisation. This generalisation gives rise to an even larger set of vector beams with more intricate polarisation distributions. Crucially, controlling the beam parameters allows engineering of vector beams with predefined polarisation trajectories on the Poincaré sphere. This capability offers potential applications, for example in optical communications, where precise polarisation control can significantly enhance data transmission and security.
\end{abstract}

\setboolean{displaycopyright}{true}

\begin{document}

\maketitle

The field of structured light has expanded significantly over the last few decades, with vector beams emerging as a prominent class due to their unique properties. Of particular interest are those in which the spatial and polarisation degrees of freedom (DoF) are coupled in a nonseparable fashion, giving rise to inhomogeneous, often exotic, polarisation patterns \cite{Rosales2018Review,Fields,Roadmap,Shen_2023,bauer2015Moebious}. Such beams have found applications across a wide range of domains, including optical trapping, microscopy, high-resolution imaging or classical and quantum communication, where precise polarisation control is essential \cite{Donato2012,Chen2013,Segawa2014,Ndagano2018,klug2023robust,rosales2024perspective,yang2021}. Traditionally, vector beams are generated in circular-cylindrical coordinates, where the spatial DoF is encoded in modes with circular symmetry and the polarisation one in the circular polarisation basis. Such beams, commonly known as cylindrical vector modes have proven valuable in various applications \cite{BergJohansen2015,Hu_2022,Liu2018,Hu2019}. However, these cylindrical symmetries impose constraints, limiting the range of achievable polarisation distributions and restricting the flexibility of polarisation manipulation.

Recent developments have explored using other spatial mode bases to achieve more intricate polarisation structures. In particular, Ince-Gauss (IG) and Mathieu-Gauss vector beams utilize elliptical spatial modes paired with circular polarisation bases, producing light beams with exotic polarisation patterns that exhibit unique symmetry properties \cite{Yao-Li2020,Rosales_2021,Zhao}. Importantly, while these vector beams allow certain control over the inhomogeneous polarisation distribution, they are inherently limited to the circular polarisation bases. As a result, the generated vector beams can not be engineered in a flexible way to contain polarisation states with predefined trajectories on the Poincaré sphere. These limitations highlight the need for a generalised framework that enables unrestricted control over polarisation distribution and orientation.


In this study, we propose a class of generalized vector modes, termed elliptical helical Ince-Gauss vector modes, where the DoF is encoded on helical Ince-Gauss modes (hIG), and the polarisation DoF is defined using an elliptical polarisation basis. This approach broadens the conventional vector beam framework, enabling an expanded set of exotic polarisation distributions. We show that the polarisation states across the entire transverse plane of each vector mode can be mapped point-to-point to the Poincar\'e sphere (PS), generating closed-loop trajectories on its surface. These trajectories can be dynamically controlled on demand, covering closed curves on the PS, enabling flexibility in polarisation manipulation. Moreover, we theoretically and experimentally characterize these beams on the High Order Poincaré Sphere (HOPS).
\begin{figure}[htp!]
    \centering
    \includegraphics[width=0.48\textwidth]{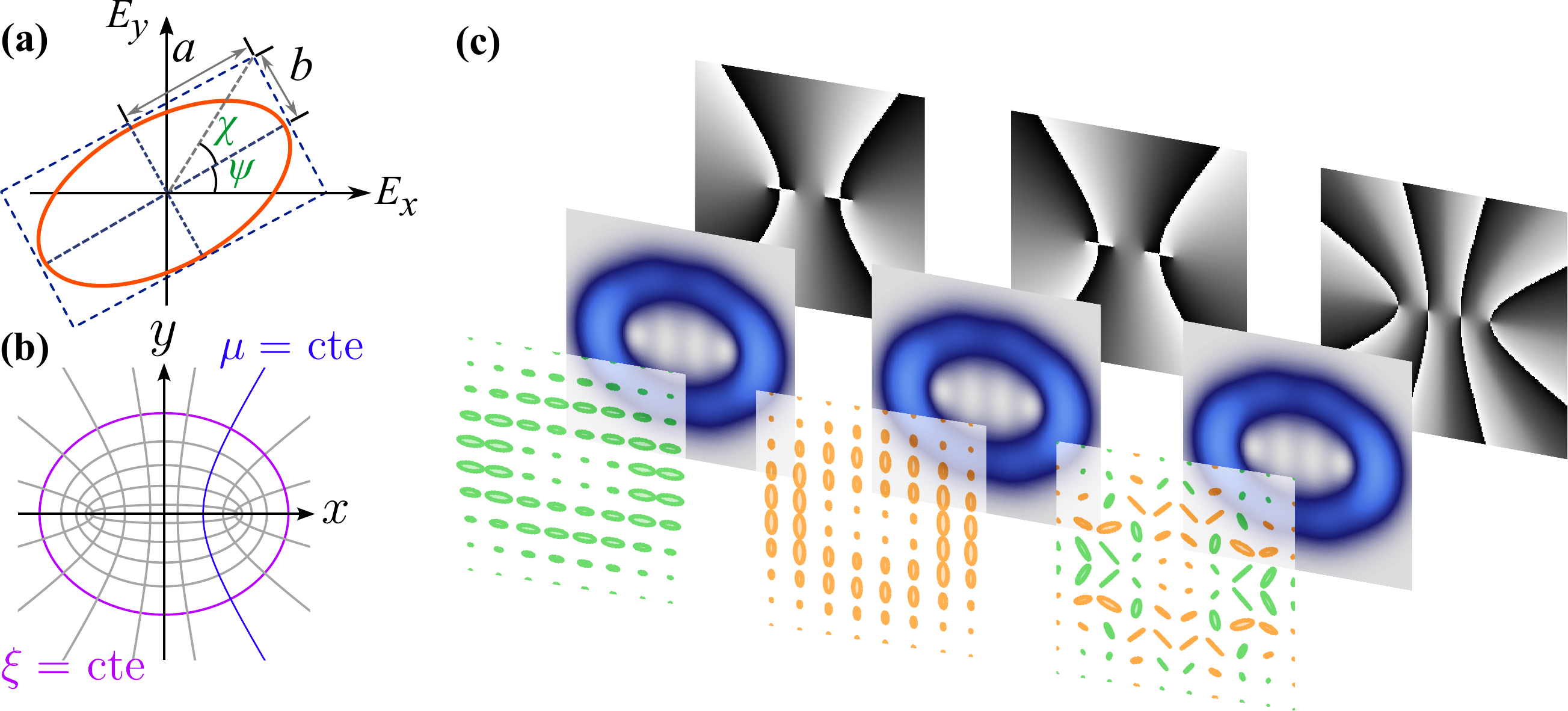}
    \caption{(a) Polarisation ellipse with its geometric characteristics determined by the amplitudes of each component and the phase difference. (b) Transverse elliptical coordinates $(\xi,\mu)$. (c) Transverse phase (back panels), intensity profiles (middle) and polarisation distribution (front) of $\text{IG}_{4,4,2}^{+}$ (left), $\text{IG}_{4,4,2}^{-}$ (middle) and $\vec{hIV}$ (right). Right and left elliptical polarisation are represented by orange and green ellipses, respectively.}
    \label{PE}
\end{figure}

As it is well-known, any polarisation state can be described mathematically by the polarisation ellipse \cite{Goldstein2011}
\begin{equation}
    \frac{E^2_x}{E^2_{0x}}+\frac{E^2_y}{E^2_{0y}}-2\frac{E_x}{E_{0x}}\frac{E_y}{E_{0y}}\cos\delta=\sin^2\delta,
    \label{Eq:PE}
\end{equation}
where $\delta$ represents the phase difference between both components of the electric field, $E_x$ and $E_y$. Equation \ref{Eq:PE} has the form of a rotated ellipse, as illustrated in Fig.\ \ref{PE}(a), where $a$ and $b$ represent the semi-major and semi-minor axes and $\psi$ the angle of the semi-major axis with respect to the horizontal axis. In addition, the ellipticity of any polarisation state is specified by the ellipticity angle $\chi$. In this way, the orientation angle and ellipticity angle of any polarisation state can be specified using the relations \cite{Goldstein2011}
\begin{equation}\label{StokesParameters}
\begin{aligned}
        \tan2&\psi=\tan2\phi\cos\delta\,, & 0\leq&\psi\leq\pi\,, \\
        \sin2&\chi=\sin2\phi\sin\delta\,, & -\pi/4\leq&\chi\leq\pi/\,,
\end{aligned}
\end{equation}
where, the parameter $\phi$ is related to the amplitudes of the electric field through the relation $\tan\phi=E_{0y}/E_{0x}$, $0\leq\phi\leq\pi/2$.  Notice that, for $-\pi/4\leq\chi< 0$ (see Eq.\ \ref{StokesParameters}) we have left-handed polarisation, whereas for $0<\chi\leq\pi/4$ right-handed.

Similarly, any set of solutions to the paraxial Helmholtz equation can be referred to as a beam mode. Commonly known examples include the Laguerre-Gaussian modes, which are solutions in polar cylindrical coordinates. In this work, we use the more general case where the paraxial Helmholtz equation is solved in elliptical cylindrical coordinates $\vec{r} = (\xi,\mu,z)$ (Fig.\ \ref{PE}(b)), which leads to the IG modes \cite{Bandres2004}
\begin{align}\label{eq:IGe}
	\text{IG}_{p,m,\varepsilon}^e({\bf r}) &= \frac{Cw_0}{w(z)}C_p^m(i\xi,\varepsilon)C_p^m(\mu,\varepsilon)\text{e}^{-\frac{\rho^2}{w(z)}}\text{e}^{-i\left(kz+Z-\Phi\right)}\,,\\ \label{eq:IGo}
    \text{IG}_{p,m,\varepsilon}^o({\bf r}) &= \frac{Sw_0}{w(z)}S_p^m(i\xi,\varepsilon)S_p^m(\mu,\varepsilon)\text{e}^{-\frac{\rho^2}{w(z)}}\text{e}^{-i\left(kz+Z-\Phi\right)}\,,\\ \label{eq:IGh}
    \text{IG}_{p,m,\varepsilon}^\pm({\bf r}) &= \text{IG}_{p,m,\varepsilon}^e({\bf r}) \,\pm\, i\, \text{IG}_{p,m,\varepsilon}^o({\bf r})\,,
\end{align}
where $C_p^m(\cdot,\varepsilon)$ and $S_p^m(\cdot,\varepsilon)$ are the even and odd Ince polynomials, respectively. $\omega(z) = \omega_0\sqrt{1 + \left(z/z_R\right)^2}$ is the beam waist at the plane $z$, and $\omega_0 = \omega(0)$. Further, $z_R=\pi\omega_0^2/\lambda$ is the Rayleigh distance, $\Phi(z)=(p+1)\arctan(z/z_R)$ is the Gouy phase and $Z(z)=kr^2/2R(z)$ with $R(z)=z+z_R^2/z$. Finally, $\rho$ is the transverse radial coordinate, $k=2\pi/\lambda$ is the wavenumber, and $C$ and $S$ are normalization constants.  The transverse coordinates $(\xi,\mu)$ are related to the Cartesian system by $x = \sqrt{2/\varepsilon \omega^2(z)} \cosh{\xi} \cos{\mu}$ and $y = \sqrt{2/\varepsilon \omega^2(z)} \sinh{\xi} \sin{\mu}$, where $\xi\in[0,\infty)$ and $\mu\in[0,2\pi)$ are the radial and angular elliptical coordinates. For even beams (Eq.\ \ref{eq:IGe}), $0\le m \le p$, for odd modes (Eq.\ \ref{eq:IGo}), $1\le m \le p$, where $p$ and $m$ must be the same. In addition to the order $p$ and degree $m$, the ellipticity parameter $\epsilon$ enables controlling the transition between the Larguerre-Gaussian (LG) beams ($\epsilon=0$) and Hermite-Gaussian (HG) beams ($\epsilon\rightarrow\infty$). The helical Ince-Gauss (Eq.\ \ref{eq:IGh}) beams are characterized by $M=[1+(p-m)/2]$ elliptic rings and $m$ physically separated vortices of unitary topological charge. 

Complex vector light modes are commonly generated as non-separable superpositions of polarisation and spatial DoF. Similarly to cylindrical vector modes, by employing the elliptical polarisation basis and helical Ince-Gaussian beams, we can represent the helical Ince-Gaussian vector beam as:
\begin{equation}
\vec{hIV}_{p,m,\varepsilon}(\vec{r}) = \cos\theta \, \text{IG}_{p,m,\epsilon}^{+}(\vec{r})\,\hatt{e}_{ER} + \sin\theta \, e^{i\alpha} \, \text{IG}_{p,m,\epsilon}^{-}(\vec{r})\,\hatt{e}_{EL}\,,
\label{HIV}
\end{equation}
where $\theta \in [0, \pi/2]$ controls the weighting factors that allow the field $\vec{hIV}$ to transition from purely scalar ($\theta=0$ and $\theta=\pi/2$) to purely vector ($\theta=\pi/4$). Besides, the term $\alpha\in[-\pi/4,\pi/4]$ represents the intermodal phase between right- and left-handed elliptical polarisation components $\hatt{e}_{ER}$ and $\hatt{e}_{EL}$, respectively.  To transition between circular and elliptical polarisation basis, we perform the following transformation \cite{Goldstein2011, chipman2018polarized}
\begin{equation}
\begin{pmatrix}
    E_{ER} \\ E_{EL}
\end{pmatrix} = 
\begin{pmatrix}
  \text{cos}(\beta/2)  &  \text{exp}(i\eta)\text{sin}(\beta/2) \\
  \text{sin}(\beta/2)  & -\text{exp}(i\eta)\text{cos}(\beta/2)
\end{pmatrix}
\begin{pmatrix}
    E_{R} \\ E_{L}
\end{pmatrix}
\end{equation}
where $E_{EL}$ and $E_{ER}$ are the elliptical components, and $E_{L}$ and $E_{R}$ are the circular components. The parameters $\beta$ and $\eta$ determine the angle and ellipticity of each elliptical state, respectively.  Figure \ref{PE}(c) shows, as an example, two scalar modes $\text{IG}_{4,4,2}^{+}$ and $\text{IG}_{4,4,2}^{-}$ with right- and left-handed elliptical polarisation with parameters $\eta=3\pi/4$ and $\beta=3\pi/2$ depicted by orange and green ellipses, respectively. The front, middle, and back panels show the polarisation distribution, intensity profile, and phase structure, respectively. The right column in Fig.\ \ref{PE}(c) illustrates the case of $\vec{hIV}_{4,4,2}$ with $\theta=\pi/4$ and $\alpha=0$. The back panel depicts the phase of the Stokes field.  Although all beams show the same intensity pattern, the polarisation and phase distribution vary.

To generate the helical Ince-Gaussian vector beams described above, we used a digital micromirror device (DMD) following the methodology we previously proposed (see \cite{Rosales2020} for details). We characterized our vector beams via two strategies: first, by mapping the polarisation states of the vector mode over the PS; and second, by mapping the vector mode over a HOPS.  The former provides a comprehensive model for visualizing all the polarisation states:  linear states located along the equator, circular polarisations at the poles, and elliptical polarisations distributed over the rest of the sphere. The latter, on the other hand, offers a geometric representation well-suited for visualizing vector beam states \cite{PhysRevLett.107.053601, Yao-Li2020}. Each vector state corresponds to a unique point $(2\alpha,2\theta)$ on the unitary sphere, where the North and South poles represent the scalar modes $\text{IG}_{p,m,\epsilon}^{+}\,\hatt{e}_{ER}$ and $\text{IG}_{p,m,\epsilon}^{-}\,\hatt{e}_{EL}$, respectively, and the equator represents pure vectorial states. The remainder of the sphere's surface corresponds to elliptically polarised vectorial states.  It is important to note that both characterisation strategies rely on the Stokes parameters. Experimentally, we can obtain the Stokes parameters via four intensity measurements \cite{Goldstein2011, Zhaobo2019}
\begin{equation}\label{Eq:Stokes}
\begin{split}
\centering
     &S_{0}=I_{0},\hspace{19mm} S_{1}=2I_{H}-S_{0},\hspace{1mm}\\
     &S_{2}=2I_{D}-S_{0},\hspace{10mm} S_{3}=2I_{R}-S_{0},
\end{split}
\end{equation}
where $I_0$ represents the total intensity, and $I_H$, $I_D$ and $I_R$ are the intensities of the horizontal, diagonal and right-handed polarisation components, respectively. These intensities were obtained using a polarizer (P), and a quarter-wave plate (QWP) and captured with a CCD camera (FLIR FL3-U3-120S3C-C from Pointgray). Specifically, $I_H$ and $I_D$ were measured by passing the beam through the polarizer at $0^\circ$ and $45^\circ$, respectively, while $I_R$ was measured by combining the QWP at $45^\circ$ with the polarizer at $90^\circ$, all angles with respect to the horizontal. For all the experimental results we set $w_0 = 0.26$ mm.

\begin{figure}[tp]
    \centering
    \includegraphics[width=0.45\textwidth]{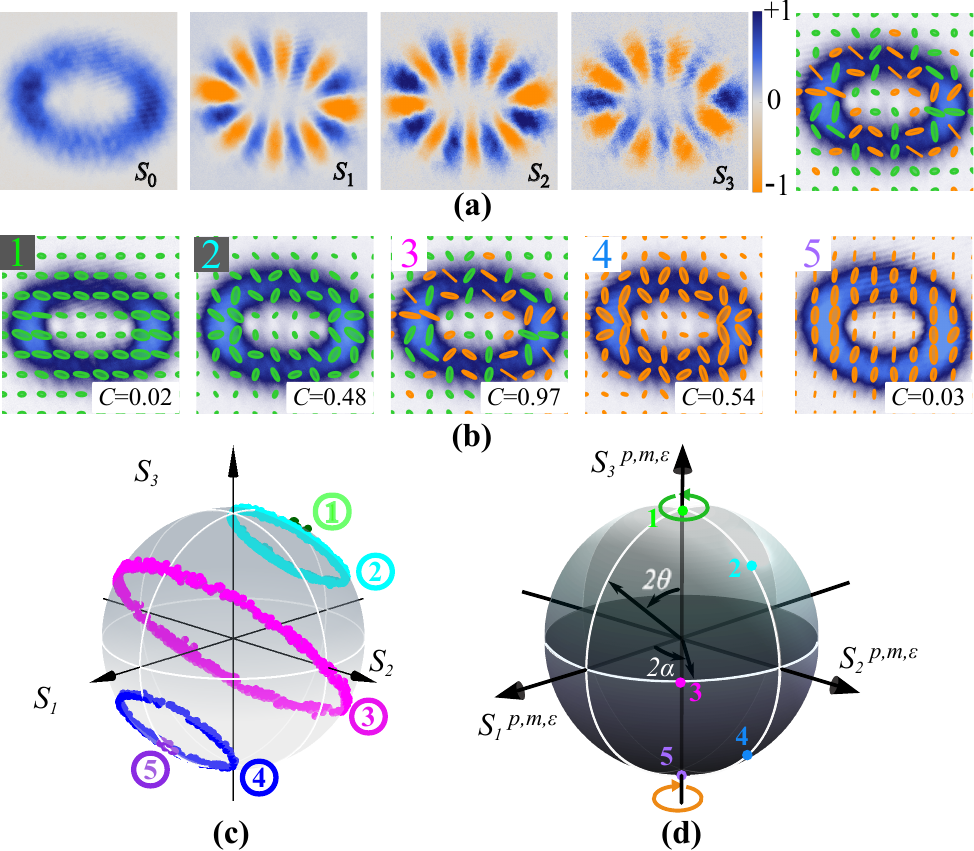}
    \caption{(a) Stokes parameters $S_0$, $S_1$, $S_2$ and $S_3$ and its reconstructed polarisation. (b) Experimental transverse intensity and polarisation distribution of the modes $\vec{hIV}_{4,4,2}$ with $\eta=3\pi/4$ and $\beta = 3\pi/2$, and parameters $\{\theta, \alpha\}$: (green, 1) $\{0, \pi/4\}$, (cyan, 2) $\{\pi/8, \pi/4\}$, (pink, 3) $\{\pi/4, \pi/8\}$, (blue, 4) $\{3\pi/8, \pi/4\}$, and (purple, 5) $\{\pi/2, \pi/4\}$. (c) Experimental trajectories of their polarisation over the PS. (d) Geometric representation on a HOPS.}
    \label{Poincare}
\end{figure}

By way of example, Fig.\ \ref{Poincare} shows a representative set of five experimental $\vec{hIV}_{4,4,2}$ vector modes generated with the polarisation ellipticity parameters $\eta=3\pi/4$ and $\beta = 3\pi/2$, for the cases $\{\theta, \alpha\}$ given by, $\{0, \pi/4\}$ (case 1 in green), $\{\pi/8, \pi/4\}$ (case 2 in cyan), $\{\pi/4, \pi/8\}$ (case 3 in pink), $\{3\pi/8, \pi/4\}$ (case 4 in blue), and $\{\pi/2, \pi/4\}$ (case 5 in purple). Here, we first show in Fig.\ \ref{Poincare}(a) an example of the measured Stokes parameters $S_{0}$, $S_{1}$, $S_{2}$ and $S_{3}$, from which the transverse polarisation distribution is reconstructed, as shown in the right panel. The reconstructed polarisation distribution for each case is shown in Fig.\ \ref{Poincare}(b). All the polarisation states contained in each vector beam are mapped over the PS, as depicted in Fig.\ \ref{Poincare}(c). Notice that all cases show closed-loop trajectories. Finally, Fig.\ \ref{Poincare}(d) shows all the vector modes mapped as unique points $\{\theta, \alpha\}$ onto the surface of HOPS.

We further quantify the non-separability through the {\it Concurrence} $(C)$, also known as Vector Quality Factor (VQF), a measure borrowed from quantum mechanics.  In particular, $C$ quantifies the degree of coupling between the spatial and polarisation DoF, and can be expressed as \cite{McLaren2015, Selyem2019}
\begin{align}\label{eq:vqf}
    C = \sqrt{1 - \left(\frac{\mathbb{S}_1}{\mathbb{S}_0} \right)^2 + \left(\frac{\mathbb{S}_2}{\mathbb{S}_0} \right)^2 + \left(\frac{\mathbb{S}_3}{\mathbb{S}_0} \right)^2},
\end{align}
where $\mathbb{S}_j = \iint_{R^2} S_j dA$, and $C$ ranges from 0 (separable) to 1 (non-separable). The corresponding $C$ values of the examples above are shown in the insets of Fig.\ \ref{Poincare}(b).  Moreover, one can show that a vector mode whose mapping over the PS corresponds to a great circle will be maximally non-separable, i.e. $C = 1$.  In order to prove this, we computed Eq.\ \ref{eq:vqf}, starting from $\vec{hIV}_{p,m,\epsilon}$ with $\theta=\pi/4$. Following a straightforward calculation led to $\mathbb{S}_j = 0$, for $j=1,2,3$, and $\mathbb{S}_0 = 1$.

\begin{figure*}[htp!]
    \centering
    \includegraphics[width=0.82\textwidth]{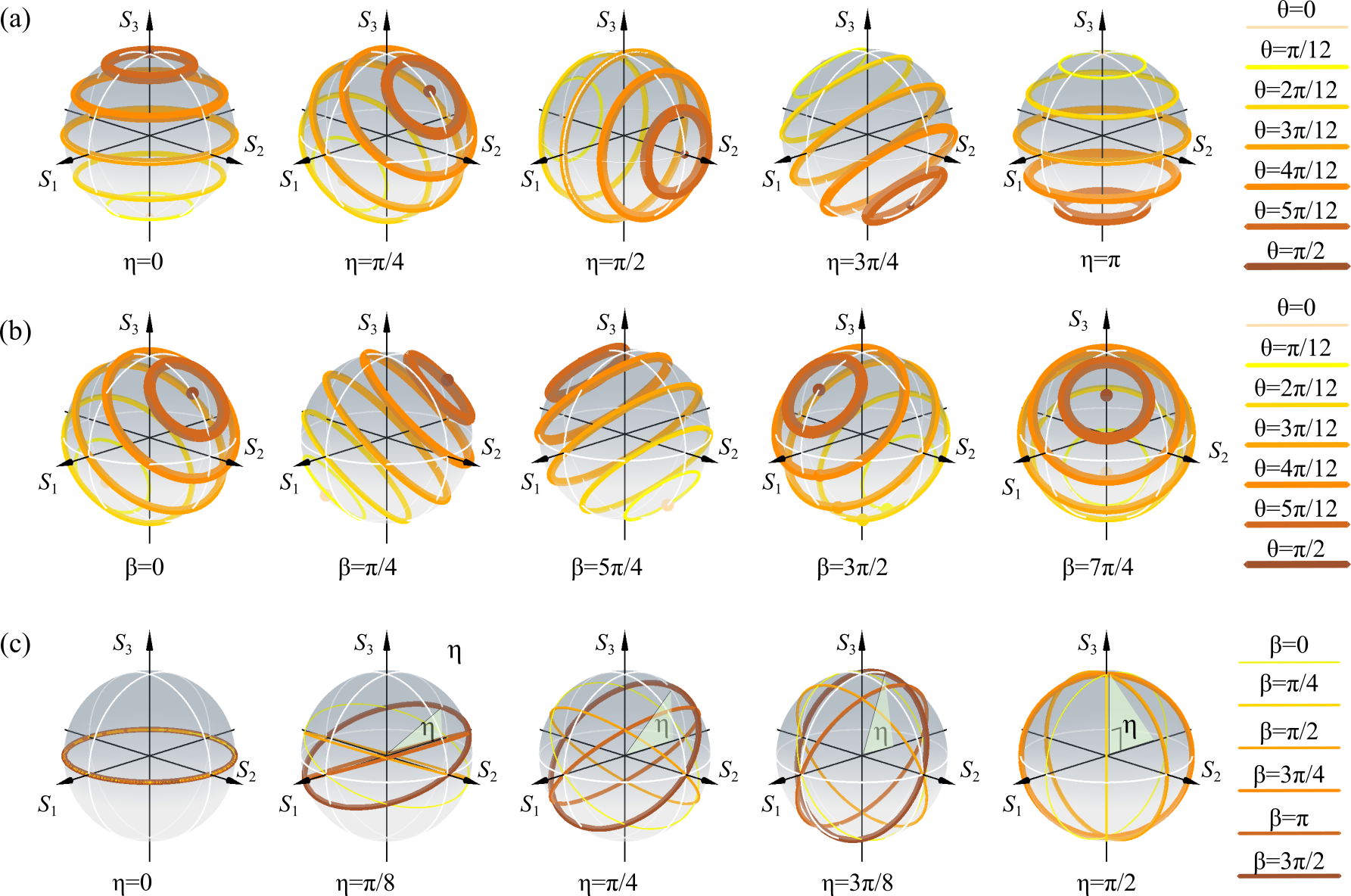}
    \caption{Polarisation trajectories on the PS as a function of (a) $\eta$ and (b) $\beta$ for the specific mode $\vec{hIV}_{4,4,2}$, with $\beta = 0$ and $\eta = \pi/4$ held constant, respectively. In both cases, different trajectories are shown corresponding to $\theta$ varying in the interval [0, $\pi$/2]. In (c) we show the polarisation trajectories on the PS as a function of $\eta$, while holding $\theta = \pi/4$ constant.  Different trajectories are shown corresponding to $\beta$ within the interval $[0, 3\pi/2]$.} 
    \label{eta&beta}
\end{figure*}

An exciting feature of the trajectories over the PS is their ability to be controlled on demand by tuning the parameters $\theta$, $\alpha$, $\eta$ and $\beta$. We simulated and explored some trajectories on the PS, as shown in Fig.\ \ref{eta&beta}. When $\eta$ increases from $0$ to $\pi$, the trajectories corresponding to some fixed values of $\theta$ rotate clockwise around the $S_1$ axis, as shown in Fig.\ \ref{eta&beta}(a). Similarly, as $\beta$ increases from $0$ to $7\pi/4$, the trajectories rotate counterclockwise around the $S_3$ axis, illustrating the influence of $\beta$ on these paths (Fig.~\ref{eta&beta}(b)). Moreover, we found that the trajectories over the PS of different vector modes can be confined within specific angular ranges from the centre of the PS. For example, in Fig. \ref{eta&beta}(c), we demonstrate polarisation trajectories for elliptical vector modes with $\beta$ ranging from $0$ to $3\pi/2$ as $\eta$ increases from $0$ to $\pi/2$. When $\eta$ is zero, \bl{all} the trajectories -for different $\beta$ values- align along the equator. As $\eta$ increases to $\pi/4$, the trajectories rotate counterclockwise with respect to varying $\beta$ values, although the maximum deviation along the $S_3$ axis is limited by $\eta$. We further illustrate this effect for $\eta=\pi/2$, where all trajectories follow lines of latitude connecting the North and South poles of the PS. 

\begin{figure}[tb]
    \centering
    \includegraphics[width=0.48\textwidth]{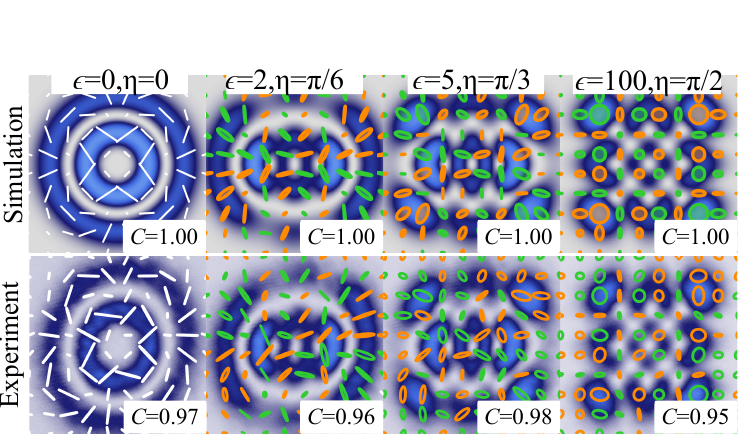}
    \caption{Transverse intensity profiles overlapped with the polarisation distribution for $\vec{hIV}_{5,3,\varepsilon}$ with $\alpha = 0$, $\beta = \pi$, and $\theta = \pi/4$. From left to right, the values of $\{\epsilon, \eta\}$ are $\{0,0\}$, $\{2,\pi/6\}$, $\{5,\pi/3\}$, and $\{100,\pi/2\}$.}
    \label{Eccentricity}
\end{figure}

So far we have only explored the ellipticity parameter $\eta$ of the polarisation DoF and demonstrated that it is possible to achieve a wide variety of polarisation distributions. We now incorporate the ellipticity parameter $\epsilon$ of the spatial DoF to achieve an even larger set of vector beams. More precisely, as we have mentioned before $\vec{hIV}_{p,m,\varepsilon}(\vec{r})$ vector beams generated in the circular polarisation basis evolve from LG to HG vector beams as $\epsilon$ increases from 0 to $\infty$ maintaining a transverse polarisation state that contains only linear states. With the parameter $\eta$ increasing from 0 to $\pi/2$, we can change the polarisation ellipticity from linear to circular, respectively. Hence, it is possible to generate $\vec{hIV}$ modes with completely different polarisation distributions. In other words, this generalised framework allows generating a $\vec{hIV}$ vector mode of parameter $\epsilon$ with linear, elliptical or circular polarisation distributions. This provides a natural basis with potential applications in, for example, quantum cryptography. To illustrate this, in Fig.\ \ref{Eccentricity} we show an example of $\vec{hIV}$ modes generated with different values of mode ellipticity $\epsilon$ and polarisation ellipticity $\eta$, simulations on top and experimental results on the bottom. For this examples we used the mode $\vec{hIV}_{5,3,\varepsilon}$, with parameters $\alpha = 0$, $\beta = \pi$, and $\theta = \pi/4$. As $\eta$ changes along with $\epsilon$, the intensity distribution transitions from circular to rectangular, while the ellipticity of the polarisation states changes (in some regions from linear to circular).


To summarise, in this manuscript, we have introduced a class of elliptical vector light modes, highlighting their ability to feature diverse polarisation distributions. By encoding the spatial DoF with helical Ince-Gauss modes and using an elliptical polarisation basis, we have demonstrated that these modes can produce a wide variety of polarisation states. Through the mapping of these states across the transverse plane, we obtain closed curve trajectories on the PS. Furthermore, we emphasize on the control of these trajectories through the adjustable parameters of the $\vec{hIV}$ modes. This capability not only adds a layer of versatility to the manipulation of light but also opens up promising avenues for practical applications in fields such as optical communication.

\begin{backmatter}
\bmsection{Funding}
This work was supported by Zhejiang Provincial Natural Science Foundation of China under Grant No. LQ23A040012.
 
\bmsection{Disclosures}
The authors declare that there are no conflicts of interest related to this article.


\bmsection{Data Availability Statement}
All the data are available from the corresponding author upon reasonable request.
\end{backmatter}


\bibliographyfullrefs{References}
\end{document}